\documentclass[conference]{IEEEtran}

\usepackage{amsmath,amssymb,amsfonts}
\usepackage{algorithmic}
\usepackage{graphicx}
\usepackage{textcomp}
\usepackage{xcolor}
\def\BibTeX{{\rm B\kern-.05em{\sc i\kern-.025em b}\kern-.08em
    T\kern-.1667em\lower.7ex\hbox{E}\kern-.125emX}}
\usepackage[sorting=none]{biblatex}
\usepackage{hyperref}
\usepackage{subcaption}
\usepackage{multirow}
\usepackage{listings}
\definecolor{codegreen}{rgb}{0,0.6,0}
\definecolor{codegray}{rgb}{0.5,0.5,0.5}
\definecolor{codepurple}{rgb}{0.58,0,0.82}
\definecolor{backcolour}{rgb}{0.95,0.95,0.95}
\lstdefinestyle{mystyle}{
    backgroundcolor=\color{backcolour},   
    commentstyle=\color{codegreen},
    keywordstyle=\color{codepurple},
    numberstyle=\tiny\color{codegray},
    stringstyle=\color{brown},
    basicstyle=\ttfamily\footnotesize,
    frame=single,
    breakatwhitespace=false,         
    breaklines=true,
    postbreak=\mbox{\textcolor{red}{$\hookrightarrow$}\space},
    captionpos=b,                    
    keepspaces=true,                 
    numbers=left,                    
    numbersep=5pt,                  
    showspaces=false,                
    showstringspaces=false,
    showtabs=false,                  
    tabsize=2,
    xleftmargin=0pt,
    xrightmargin=0pt
}
\begin{document}

\title{Circuit Partitioning and Full Circuit Execution: A Comparative Study of GPU-Based Quantum Circuit Simulation}


\author{
\IEEEauthorblockN{Kartikey Sarode}
\IEEEauthorblockA{\textit{Department of Computer Science} \\
\textit{San Francisco State University}\\
San Francisco, USA \\
ksarode@sfsu.edu}
\and
\IEEEauthorblockN{Daniel E. Huang}
\IEEEauthorblockA{
\textit{San Francisco State University}\\
San Francisco, CA, USA \\
\href{mailto:danehuang@sfsu.edu}{danehuang@sfsu.edu}}
\and
\IEEEauthorblockN{E. Wes Bethel}
\IEEEauthorblockA{\textit{San Francisco State University} \\
San Francisco, CA, USA \\
\textit{Lawrence Berkeley National Laboratory}\\
Berkeley, CA, USA \\
\href{mailto:ewbethel@sfsu.edu}{ewbethel@sfsu.edu}}
}

\maketitle

\thispagestyle{plain}
\pagestyle{plain}

\begin{abstract}


Executing large quantum circuits is not feasible using the currently available NISQ (noisy intermediate-scale quantum) devices. The high costs of using real quantum devices make it further challenging to research and develop quantum algorithms. As a result, performing classical simulations is usually the preferred method for researching and validating large-scale quantum algorithms. However, these simulations require a huge amount of resources, as each additional qubit exponentially increases the computational space required.

Distributed Quantum Computing (DQC) is a promising alternative to reduce the resources required for simulating large quantum algorithms at the cost of increased runtime. This study presents a comparative analysis of two simulation methods: circuit-splitting and full-circuit execution using distributed memory, each having a different type of overhead.

The first method, using CutQC, cuts the circuit into smaller subcircuits and allows us to simulate a large quantum circuit on smaller machines. The second method, using Qiskit-Aer-GPU, distributes the computational space across a distributed memory system to simulate the entire quantum circuit. Results indicate that full-circuit executions are faster than circuit-splitting for simulations performed on a single node. However, circuit-splitting simulations show promising results in specific scenarios as the number of qubits is scaled.

\end{abstract}


\section{Introduction}
\label{chap:introduction}

The current NISQ-era (Noisy Intermediate-Scale Quantum era) quantum computers are sensitive to their environment (noisy), prone to quantum decoherence and are not large enough (considering the number of qubits) to achieve quantum advantage. Many hybrid algorithms (where some calculations are offloaded to a classical processor) designed for these computers show promising results in solving problems that are too big or take an exponential amount of time for classical computers. However, large problems still cannot be executed on these machines due to the low number of qubits. To overcome this limitation, Distributed Quantum Computing (DQC) can be used, where the quantum circuit for a specific problem can be split and divided across multiple quantum nodes for execution. Each node has a subset of qubits assigned to it, and non-local CNOT gates (where the control and target qubits lie on different nodes) are implemented by moving qubits between quantum nodes. However, to achieve this, we require a reliable quantum communication channel to let us move qubits from one node to another without losing information. This is still an active area of research in the scientific community.  

Considering all the factors mentioned above, it takes a lot of effort to successfully develop and validate quantum algorithms for solving large problems (requiring hundreds of qubits). To tackle this problem of developing and validating algorithms that require large quantum circuits, a lot of work has been done to be able to simulate large quantum circuits on classical machines in a distributed manner. However, different simulation methods have various kinds of overhead associated with them, which makes it hard to decide which approach should be followed while simulating a large quantum circuit. This work aims to comprehensively analyse two different approaches to simulating large quantum circuits, i.e., circuit-splitting versus full-circuit execution using distributed memory. The former aims to split the quantum circuit into multiple subcircuits to reduce the width of circuits to simulate and allow the simulation of large quantum circuits using smaller machines. The latter approach simulates the entire circuit by distributing the statevector of the quantum circuit across a distributed memory space. More information about both approaches can be found in Section \ref{sec:circuit-cutting} and Section \ref{sec:gpu-simulation}.  

Results show that the full-circuit simulation method is faster than the circuit-splitting method. However, it is not straightforward to conclude which simulation method should be used for simulating large-scale quantum circuits since many parameters must be evaluated and compared to select a feasible simulation method. This study aims to list the pros and cons of each simulation method and shed more light on what parameters can help decide which simulation method is best to use.
\section{Literature Review}
\label{chap:lit-review}

\vspace{\baselineskip}

\subsection{Origins of DQC}

While the idea of quantum computing has been around for a couple of decades, research related to Distributed Quantum Computing (DQC) started to gain traction at the start of the century. Initial works focused on developing theoretical methods for distributed computing by sharing entangled qubits between quantum devices. Yimsiriwattana, A. et al. \cite{yimsiriwattana2004generalized} presented two primitive operations \emph{cat-entangler} and \emph{cat-disentangler} that can be used to create non-local CNOT gates (which is one of the universal gates in quantum computing). The concept worked on the assumption that an entangled qubit pair is established between two quantum devices. The \emph{cat-entangler} can then transform a control qubit $\alpha|0\rangle + \beta|1\rangle$ and the entangled pair $\frac{1}{\sqrt{2}}(|00\rangle + |11\rangle)$ into a "cat-like" state $\alpha|00\rangle+\beta|11\rangle$. This state then allows both the computers to share the control qubit. In the same year, another work by Yimsiriwattana, A. et al. \cite{yimsiriwattana2004distributed} utilized these non-local CNOT gates to propose a distributed version of the popular Shor's algorithm. \par

\subsection{Shift towards optimizing non-local operations}

More recently, the focus has shifted towards developing frameworks which minimize the non-local communication required in distributed versions of popular quantum algorithms. Initial frameworks used graph partitioning algorithms to find partitions of qubits to assign to different quantum computers to minimize the overall non-local communication overhead. Baker, J. et al. \cite{baker2020time} proposed a new circuit mapping scheme for cluster-based machines called FPG-rOEE that reduces the number of operations added for non-local communication across various quantum algorithms with the help of a dynamic partitioning scheme. Wu, A. et al. \cite{wu2022collcomm} proposes a compiler framework, CollComm, to optimize the collective communication in distributed quantum programs. The framework achieves this by decoupling inter-node operations from communication qubits. Recent work by Sundaram, R. et al. \cite{sundaram2023distributing} proposes two algorithms i.e. LocalBest and ZeroStitching to distribute a quantum circuit across a network of heterogeneous quantum computers to minimize the number of teleportations required to implement non-local gates. \par

\subsection{Circuit splitting}

Apart from the above works that focus on minimizing the number of non-local operations, a lot of work is also being done to eliminate the non-local operations completely. These works have a common theme where the aim is to split/cut the quantum circuit into multiple sub-circuits that only have local gates. Classical post-processing is then used on the outputs of these sub-circuits to construct the output of the original un-cut circuit. Peng, T. et al. \cite{peng2020simulating} propose a cluster simulation scheme to represent a quantum circuit as a tensor network and then partition the tensor networks into multiple smaller networks. Edges between networks are simulated by decomposing the interaction based on the observables and states from the computational basis. Chen, D. et al. \cite{chen2023efficient} adopt a similar approach where they cut a qubit wire to split the circuit into sub-circuits. Intuitively, cutting a qubit wire can be thought of as classically passing information of a quantum state along each element in a basis set. Their work exploits this and proposes that certain circuits have a \emph{golden cutting point}, which enjoys a reduction in measurement complexity and classical post-processing cost. However, identifying these \emph{golden cutting points} is non-trivial and a circuit is not guaranteed to have one. \par

Work done by Tang, W. et al. \cite{tang2021cutqc} proposes CutQC, a scalable hybrid computing approach to cutting circuits. Using a mixed-integer programming solver, the framework automatically locates cut locations along qubit wires in the circuit. Sub-circuits generated from these cuts can be executed independently in parallel to speed up overall execution time. The outputs from sub-circuits undergo classical post-processing to construct the output of the original un-cut circuit. This clean cutting of a circuit, however, comes at an exponential cost ($4^{K}$) in terms of the number of cuts (\emph{K}) made. A recent work by Piveteau, C. et al. \cite{piveteau2023circuit} proposes a method of circuit knitting based on quasi-probability simulation of non-local gates using operations that act locally on subcircuits. This approach makes it possible to cut CNOT gates between subcircuits, thus allowing more flexibility when choosing the cut locations. This added flexibility comes at an even greater cost though ($9^{n}$) where (\emph{n}) is the number of CNOT gates cut. They also propose a possible reduction of post-processing cost to ($4^{n}$) if classical communication is available between the quantum computers. \par

In our work here, we will be using the CutQC framework to split our circuits as it has the least amount of restrictions on the type of circuits that can be cut, and it also allows us to construct the full probability distribution of the un-cut circuit, which can be used to verify the outputs. Refer to section \ref{sec:circuit-cutting} for detailed information on how the circuit cutting works in the CutQC framework. Our work builds on the CutQC framework by enabling GPU support for circuit simulations. Our work mainly aims to compare the performance of both approaches, i.e., circuit-splitting and full-circuit simulations, when GPUs are used for simulating the circuits and gain more insights into the pros and cons of both approaches.

\section{Background}
\label{chap:background}

This section contains detailed information on some of the topics required to completely understand the tools and software used in this study. Section \ref{sec:circuit-cutting} describes the circuit-cutting framework CutQC and how it works, whereas Section \ref{sec:gpu-simulation} talks about how GPU-based simulation works in Qiskit.

\subsection{Circuit Cutting using CutQC}
\label{sec:circuit-cutting}

The CutQC framework cuts a quantum circuit along its qubit wires (vertical cuts). The vertical cuts use the fact that a unitary matrix of any arbitrary operation can be decomposed into any set of orthonormal matrix bases. To demonstrate this, the CutQC paper by Tang, W. et al. \cite{tang2021cutqc} uses the basis composed of the set of Pauli matrices \emph{I}, \emph{X}, \emph{Y} and \emph{Z}. Consider an arbitrary 2$\times$2 matrix \textbf{A} expressed as a linear combination of the Pauli bases, $\textbf{A} = \alpha\emph{X} + \beta\emph{Y} + \gamma\emph{Z} + \delta\emph{I}$. The trace of this matrix with a Pauli matrix, for example, with \emph{X} is 
$$ \emph{Tr} \left( \textbf{A}\emph{X} \right) =  \alpha\emph{Tr} \left( X^{2} \right) + \beta\emph{Tr} \left( X\emph{Y} \right) + \gamma\emph{Tr} \left( X\emph{Z} \right) + \delta\emph{Tr} \left( X \right) = 2\alpha$$
This is using the fact that Pauli matrices are traceless (resulting in $\emph{Tr} \left( X \right)=0$), they anti-commute, which implies their product has zero trace (resulting in $\emph{Tr} \left( X\emph{Y} \right)=\emph{Tr} \left( X\emph{Z} \right)=0$) and the square of a Pauli matrix is the identity matrix (resulting in $\emph{Tr} \left( X^{2} \right)=\emph{Tr} \left( I \right)=2$). Similarly, we can show that $\emph{Tr} \left( \textbf{A}\emph{Y} \right) = 2\beta$, $\emph{Tr} \left( \textbf{A}\emph{Z} \right) = 2\gamma$ and $\emph{Tr} \left( \textbf{A}\emph{I} \right) = 2\delta$.  

\noindent Using the above, we can decompose matrix A as 
\begin{equation} \label{eq:3.1}
    \textbf{A} = \frac{ \emph{Tr} \left( \textbf{A}\emph{I} \right) I + \emph{Tr} \left( \textbf{A}\emph{X} \right) X + \emph{Tr} \left( \textbf{A}\emph{Y} \right) Y + \emph{Tr} \left( \textbf{A}\emph{Z} \right) Z }{2}
\end{equation}

However, Eq. \ref{eq:3.1} requires access to complex amplitudes, which are unavailable on quantum computers. To execute this on a quantum computer, we need to decompose the Pauli matrices further into their eigenbasis and organize the terms.  

We obtain the following identity after this decomposition.
\begin{equation} \label{eq:3.2}
    \textbf{A} = \frac{A_1 + A_2 + A_3 + A_4}{2}
\end{equation}
where
\begin{equation} \label{eq:3.3}
    A_1 = [ \emph{Tr} \left( \textbf{A} \emph{I} \right) + \emph{Tr} \left( A\emph{Z} \right) ] |0\rangle\langle0|
\end{equation}
\begin{equation} \label{eq:3.4}
    A_2 = [ \emph{Tr} \left( \textbf{A} \emph{I} \right) - \emph{Tr} \left( A\emph{Z} \right) ] |1\rangle\langle1|
\end{equation}
\begin{equation} \label{eq:3.5}
    A_3 = \emph{Tr} \left( \textbf{A} \emph{X} \right) [ 2 |\text{+}\rangle\langle\text{+}| - |0\rangle\langle0| - |1\rangle\langle1| ]
\end{equation}
\begin{equation} \label{eq:3.6}
    A_4 = \emph{Tr} \left( \textbf{A} \emph{Y} \right) [ 2 |\text{+}i\rangle\langle\text{+}i| - |0\rangle\langle0| - |1\rangle\langle1| ]
\end{equation}

\noindent This decomposition can be achieved if we use the below equations
$$ \emph{I} = |0\rangle\langle0| + |1\rangle\langle1| $$
$$ \emph{Z} = |0\rangle\langle0| - |1\rangle\langle1| $$
$$ \emph{X} = 2 |\text{+}\rangle\langle\text{+}| - \emph{I} $$
$$ \emph{Y} = 2 |\text{+}i\rangle\langle\text{+}i| - \emph{I} $$

\noindent The first two terms can be substituted in Eq. \ref{eq:3.1} and grouped together to get Eqs. \ref{eq:3.3} and \ref{eq:3.4} whereas the last two terms help get Eqs. \ref{eq:3.5} and \ref{eq:3.6}.

\begin{figure}[htbp]
    \centering
    \includegraphics[alt={Circuit edge decomposition when a cut is made to a quantum circuit. We get three different upstream measurement circuits and four different downstream initialization circuits}, width=\linewidth]{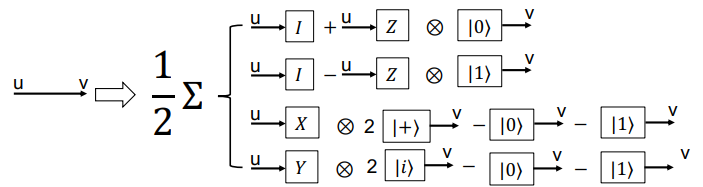}
    \caption{Procedure to cut one qubit wire. The wire between vertices $u$ and $v$ (left) can be cut by (as shown on the right) summing over four pairs of measurement circuits appended to $u$ and state initialization circuits prepended to $v$. Measurement circuits in the \emph{I} and \emph{Z} basis have the same physical implementation. The three different upstream measurement circuits and the four different downstream initialization circuits are now separate and can be independently evaluated. (Image Source: Tang, W. et al. \cite{tang2021cutqc}) }
    \label{fig:edge-decomposition}
\end{figure}

Each trace operator $A_{i}$ corresponds physically to measure the qubit in one of the Pauli bases, and each density matrix corresponds physically to initialize the qubit in one of the eigenstates. Figure \ref{fig:edge-decomposition} shows the resulting subcircuits and the reconstruction procedure incurred when making a single cut. Since measuring a qubit in the $I$ and $Z$ basis corresponds to the output, we have three upstream circuits (measuring in $I$, $X$ and $Y$ bases) and four downstream circuits (initializing with $|\text{0}\rangle$, $|\text{1}\rangle$, $|\text{+}\rangle$ and $|\text{-}\rangle$).

\begin{figure}[htbp]
    \centering
    \includegraphics[alt={Example of cutting a five-qubit circuit into two smaller subcircuits of three qubits each using the CutQC framework.}, width=\linewidth]{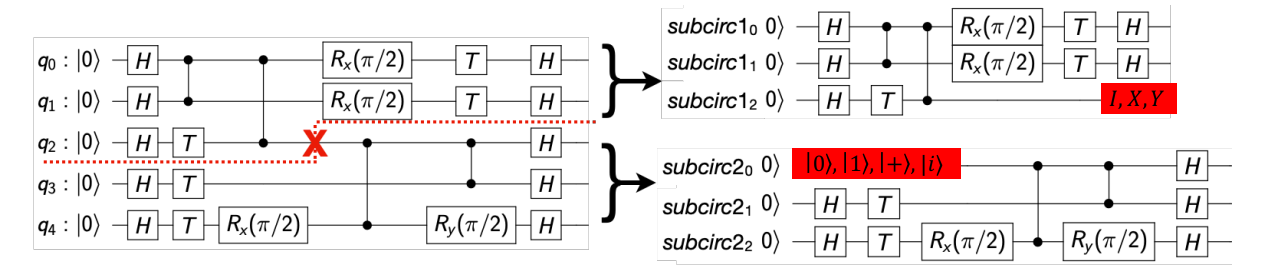}
    \caption{Example of cutting a five-qubit circuit into two smaller subcircuits of three qubits each. The subcircuits are produced by cutting the $q_2$ wire between the first two $cZ$ gates. The three variations of $subcircuit_1$ and four variations of $subcircuit_2$ can then be evaluated on a 3-qubit quantum device, instead of a 5-qubit device. The classical postprocessing involves summing over four Kronecker products between the two subcircuits for the one cut made. (Image Source: Tang, W. et al. \cite{tang2021cutqc}) }
    \label{fig:cut-example}
\end{figure}

Figure \ref{fig:cut-example} shows an example of how the cut process looks like using a sample circuit of 5 qubits. The cut splits the circuit into two subcircuits with 3 qubits each. Note that the qubit \textit{subcirc}$1_2$ does not appear in the final output of the uncut circuit. Therefore, each shot obtained from executing subcircuit 1 needs to be multiplied by $\pm$1 depending on the measurement outcome of \textit{subcirc}$1_2$. This is called 'shot attribution' and contributes to the final output as

\begin{equation} \label{eq:3.7}
\begin{split}
    \begin{cases}
        & \overline{xx0},\overline{xx1} \rightarrow +\overline{xx} \quad M_2 = I, \\
        & \overline{xx0} \rightarrow +\overline{xx} \\
        & \overline{xx1} \rightarrow -\overline{xx} \quad \quad \quad \text{otherwise}
    \end{cases}
\end{split}
\end{equation}
where $\overline{xx}$ is the measured outcome of qubits \textit{subcirc}$1_0$ and \textit{subcirc}$1_1$.

Using the above equations, the probability of a particular state, e.g. $|01010\rangle$, can be calculated for the uncut circuit in Figure \ref{fig:cut-example}. Only the first two qubits from subcircuit 1 contribute to the final state (i.e. $|01\rangle$), so we must attribute the shots for this subcircuit. Using Eq. \ref{eq:3.7} and Eqs. \ref{eq:3.3}...\ref{eq:3.6}, we get the reconstruction terms as 
\begin{equation*}
\begin{split}
    & p_{1,1} = p(|010\rangle | I) + p(|011\rangle | I) + p(|010\rangle | Z) - p(|011\rangle | Z) \\
    & p_{1,2} = p(|010\rangle | I) + p(|011\rangle | I) - p(|010\rangle | Z) + p(|011\rangle | Z) \\
    & p_{1,3} = p(|010\rangle | X) - p(|011\rangle | X) \\
    & p_{1,4} = p(|010\rangle | Y) - p(|011\rangle | Y)
\end{split}
\end{equation*}

\noindent Similarly, the state contributed from subcircuit 2 is $|010\rangle$ and the reconstruction terms for it are
\begin{equation*}
\begin{split}
    & p_{2,1} = p(|010\rangle\ |\ |0\rangle) \\
    & p_{2,2} = p(|010\rangle\ |\ |1\rangle) \\
    & p_{2,3} = 2p(|010\rangle\ |\ |+\rangle) - p(|010\rangle\ |\ |0\rangle) - p(|010\rangle\ |\ |1\rangle) \\
    & p_{2,4} = 2p(|010\rangle\ |\ |i\rangle) - p(|010\rangle\ |\ |0\rangle) - p(|010\rangle\ |\ |1\rangle)
\end{split}
\end{equation*}

\noindent The final reconstructed probability of the uncut state $|01010\rangle$ can then be calculated by summing over the four pairs of Kronecker products as shown in Figure \ref{fig:edge-decomposition} using
\begin{equation*}
    p(|01010\rangle) = \frac{1}{2} \sum^{4}_{i=1} p_{i,1} \otimes p_{2,i}
\end{equation*}

\noindent In the classical postprocessing step of CutQC, the above step is performed for all possible states in the uncut circuit.

\subsection{GPU-based Quantum Circuit simulation}
\label{sec:gpu-simulation}

In this study, the classical simulator used is the \emph{Qiskit Aer} \cite{qiskitAerGPU} simulator, which supports using GPUs to simulate the statevector of a quantum circuit. It also supports using multiple GPUs (in both single-node and multi-node configurations) to parallelise simulations based on the work done by Doi, J. et al. \cite{qiskitGPUpaper}. This section aims to provide more insight into their work.

On a classical machine, to simulate a \emph{n}-qubit statevector, we need to store all the $2^{n}$ probability amplitudes as single or double precision complex numbers. Storing the statevector for circuits using a few qubits is quite easy, but each added qubit doubles the amount of memory required to store the statevector. This makes it unfeasible to simulate circuits with more than 35 qubits (which requires more than 500 GB of RAM to store the entire statevector). To store such large statevectors, Distributed Memory is needed where probability amplitudes are split into chunks and distributed across nodes. This, however, has the drawback of increased communication overhead as amplitudes need to be exchanged across memory spaces. Minimizing data exchanges between memory spaces is a key factor in scaling the performance of a distributed memory environment. The technique developed by Doi, J. et al. reduces the amount of data exchanges by moving all the gates associated with a smaller number of qubits by inserting noiseless swap gates. These operations resemble cache blocking on a classical computer \cite{10.5555/2568134}, and the concept of optimization is similar to one on a classical computer since gates are blocked on the qubits to be accessed faster.  

\begin{figure*}[htbp]
\centering
\begin{subfigure}{0.4\textwidth}
  \centering
  \includegraphics[alt={Naive implementation of probability amplitude exchange between distributed memory spaces. This implementation requires twice the memory space.}, width=0.6\textwidth]{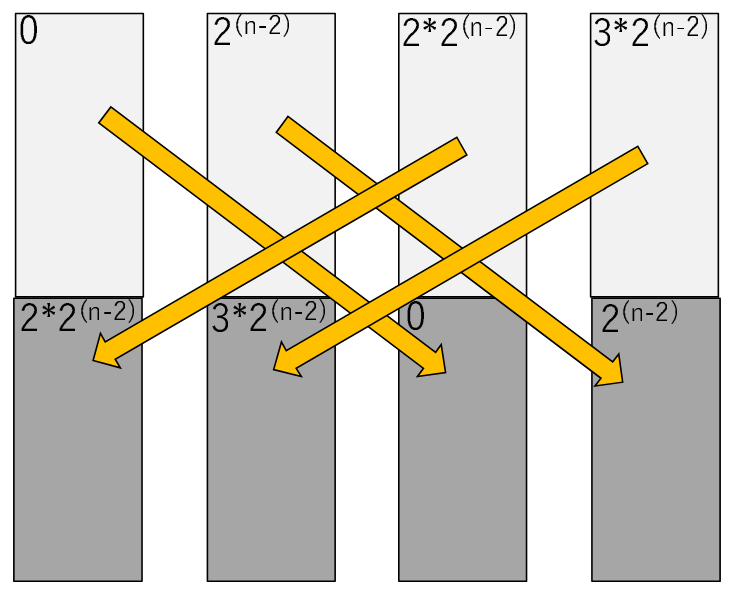}
  \caption{}
  \label{fig:naive-chunks}
\end{subfigure}%
\begin{subfigure}{0.4\textwidth}
  \centering
  \includegraphics[alt={Efficient approach of probability amplitude exchange between distributed memory spaces. The data is split into $2^{nc}$ chunks to reduce the buffer space required to exchange chunks between memory spaces.}, width=\textwidth]{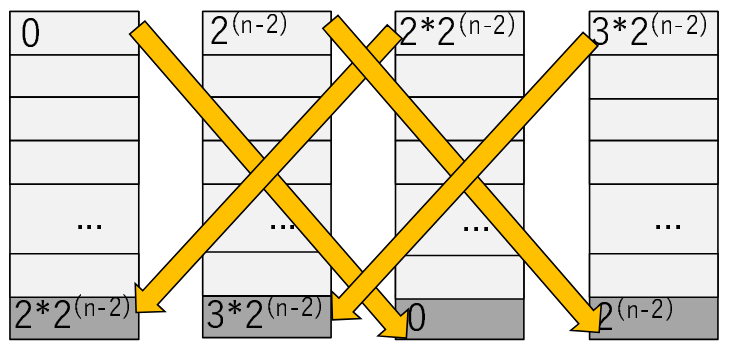}
  \caption{}
  \label{fig:efficient-chunks}
\end{subfigure}
\caption{ Different chunking methods. (a) A naive implementation of probability amplitude exchange between distributed memory spaces. In this example, we apply a gate to qubit $k=n-1$. This implementation requires twice the memory space. (b) Dividing a state vector into small chunks and performing probability amplitude exchange by chunks. So, we only need one additional chunk per memory space to perform probability amplitude exchange. (Image Source: Doi, J. et al. \cite{qiskitGPUpaper}) }
\label{fig:chunking}
\end{figure*}

The first step is to split the statevector into chunks residing in distributed memory spaces. A naive way to do this for a $n$-qubit statevector is shown in Figure \ref{fig:naive-chunks}. Here the statevector is split and spread across four distributed memory spaces with four chunks of $2^{n-2}$ probability amplitudes. In this scenario, gate operations on qubits from $0$ to $n-3$ are performed within a process, whereas data exchange is necessary for operations on $n-2$ and $n-1$ qubits. Every memory space requires a buffer for these data exchanges (shown as the darker shaded parts in the figure) to receive a copy of probability amplitudes. This approach requires double the memory space to have a copy of remote memory, which is not memory efficient. A better approach proposed in the paper is to divide the statevector into $2^{nc}$ small chunks and exchange them in a pipeline to minimize space for buffering. Figure \ref{fig:efficient-chunks} shows an example of using only one buffer to exchange chunks.

\begin{figure*}[htbp]
\centering
\begin{subfigure}{0.35\textwidth}
  \centering
  \includegraphics[alt={Example input quantum circuit having gate operations on all qubits. Data exchange is necessary to perform all gate operations.}, width=\linewidth]{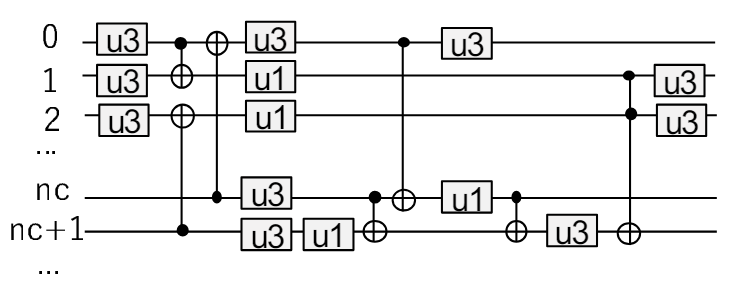}
  \caption{}
  \label{fig:gpu-input-circuit}
\end{subfigure}%
\begin{subfigure}{0.55\textwidth}
  \centering
  \includegraphics[alt={Example output quantum circuit after swap gates are inserted. All of the gate operations are now shifted to qubits less than \textit{nc} to eliminate data exchanges.}, width=\linewidth]{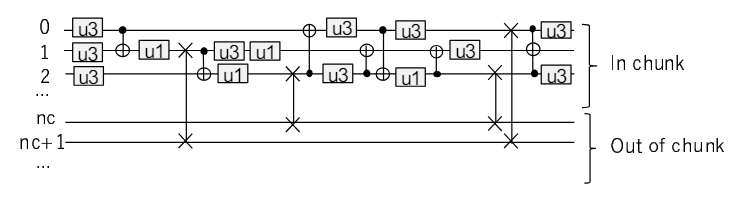}
  \caption{}
  \label{fig:gpu-transpiled-circuit}
\end{subfigure}
\caption{ Example of using cache blocking on a quantum circuit. (a) shows the input circuit consisting of u1, u3 and CNOT gates. \textit{nc} denotes the number of qubits of a chunk. The gates on qubits $>$ nc need to refer to probability amplitudes over multiple chunks to be simulated. (b) shows the output circuit after cache blocking is performed. Four swap gates are added to move all the gates to qubits $<$ \textit{nc}; now, all the gates can be performed without referring to probability amplitudes over chunks. (Image Source: Doi, J. et al. \cite{qiskitGPUpaper}) }
\label{fig:gpu-circuit}
\end{figure*}

The second step is to ensure that most gate operations are performed locally within a memory space. In other words, maximizing consecutive gate operations which act on qubits smaller than the number of chunks $nc$ since all gate operations on such qubits would be performed locally without needing any data exchange. This is done before executing the circuit in the circuit transpile phase. One of the methods is to use \textit{Qubit reordering}, where if some qubits are not updated by any of the gates, then we map them to qubits larger than or equal to $nc$ and other qubits to smaller qubits than $nc$ in simulation. This method can help to reduce data movements but cannot remove all data movements if a circuit updates $nc$ or larger qubits. In this scenario, data movement is reduced by inserting noiseless swap gates in the circuit to move as many qubits as possible to less than $nc$. An example of this can be seen in Figures \ref{fig:gpu-input-circuit} and \ref{fig:gpu-transpiled-circuit}. Figure \ref{fig:gpu-input-circuit} shows a circuit in which all qubits are updated. Here, gate operations on qubits greater than $nc$ need to refer to probability amplitudes from different chunks to be simulated. Figure \ref{fig:gpu-transpiled-circuit} shows the circuit after the cache blocking (inserting swap gates) is applied during the transpile phase. Here, using four swap gates, all the gate operations are moved to be applied on qubits less than $nc$, which does not require any data exchanges between chunks. Note that the execution order of some of the gates is also swapped so that all operations can be performed using a single chunk.

\section{Implementation}
\label{chap:implementation}

This section presents the work done as a part of this study along with the details of the CutQC framework. This is necessary as this study uses the CutQC framework by modifying it to perform the required benchmarks. Section \ref{sec:impl/cutqc} describes how the CutQC framework is used in this study along with the modifications done to the framework. Section \ref{sec:impl/benchmarks} provides more information on the benchmarking scripts used.

\subsection{CutQC Framework with updates}
\label{sec:impl/cutqc}

While Section \ref{sec:circuit-cutting} provides more information related to the math behind circuit cutting, this section aims to provide more insight into the CutQC framework from a programmatic perspective. The framework has three main modules i.e. \textit{Cutter}, \textit{Evaluator} and, \textit{Builder and Verifier}. To keep the benchmarks as close to the source material as possible, only the \textit{Evaluator} module is modified in this study to support evaluations on GPUs. In contrast, the other modules are kept the same except for changes required to support the most recent version of the tools in the quantum software environment, specifically Qiskit v1.0.2 vs v0.19.0 used in the original CutQC paper.

\subsubsection{Cutter}

As the name suggests, this module is used to identify suitable cut locations on a given quantum circuit and generate the subcircuit instances after the cuts have been made. The framework uses mixed-integer programming to automate the identification of cut locations to minimize the resulting classical postprocessing cost. This is done using the Gurobi optimizer \cite{gurobi}, a commercial optimizer solution (free to use for academic purposes).  


As mentioned above, no change is made to the Cutter module as a part of this study. Only changes necessary to support the latest version of Qiskit were made in the form of small updates to the parts of the module using deprecated Qiskit functions and attributes. Listing \ref{listing:impl/example-update} shows an example of such changes. To preserve brevity, all such changes are not included in this report.

\vspace{\baselineskip}
\begin{lstlisting}[caption={ Example for changes made to ensure compatibility with the latest Qiskit version 1.0.2. The original line of code on Line 3 is commented and the updated line of code is shown on Line 5. The code checks if both vertices of an edge are ``Gate operations" and adds them to a list if \textit{true}. },
label={listing:impl/example-update},
language=Python,
float]
for u, v, _ in dag.edges():
    # Commented original line of code
    # if u.type == "op" and v.type == "op":
    # Updated line of code
    if "DAGOpNode" in str(type(u)) and "DAGOpNode" in str(type(v)):
        u_id = vertex_ids[id(u)]
        v_id = vertex_ids[id(v)]
        edges.append((u_id, v_id))
\end{lstlisting}

\subsubsection{Evaluator}

This module simulates the subcircuit instances generated after cutting the circuit. Each subcircuit has to be simulated multiple times depending on the number of cuts made to the original circuit. The module spawns multiple processes to simulate the required subcircuits in parallel. The number of processes spawned depends on the size of the subcircuits simulated and the total available RAM on the machine running the module.  

Each subcircuit instance is first modified to add initialization gates/measurements depending on the combination of the Qubit Initializations and Qubit Measurements assigned to it. After this, the subcircuits are simulated using the Qiskit Aer simulator and the state probabilities are stored on disk by every process. Once all subcircuits have been evaluated and probabilities are stored on disk, the next step is executed i.e. the attribution of shots. This step adjusts the probability counts for subcircuits having qubits that do not contribute to the final uncut state. This is a crucial step required to reconstruct the probability distribution of the original uncut circuit.  

This module has been updated to support circuit evaluations on GPU using the latest version of the Qiskit Aer simulator. Listing \ref{listing:impl/eval-cpu} shows a trimmed version of the original function used to simulate circuits on CPUs using the \textit{statevector\_simulator} and Listing \ref{listing:impl/eval-gpu} shows the updated function supporting evaluation on GPUs using the \textit{statevector\_simulator}. A `device' option is added to the function to enable selecting the target device at runtime. The GPU simulator in Qiskit Aer requires additional options to be set when running a job (\textit{blocking\_enable} and \textit{blocking\_qubits}). The \textit{blocking\_qubits} option is used to set the chunk size for each memory space and needs to be tuned depending on the amount of GPU memory available on the target machine. For NVIDIA A100 GPUs on Perlmutter, the value is set to 30.

\vspace{\baselineskip}
\begin{lstlisting}[caption={ Original function to evaluate circuits using the \textit{statevector\_simulator} method on CPUs. (Source: CutQC Framework \cite{cutqc}) },
label={listing:impl/eval-cpu},
language=Python,
float]
def evaluate_circ(circuit, backend):
    circuit = copy.deepcopy(circuit)
    if backend == "statevector_simulator":
        simulator = aer.Aer.get_backend("statevector_simulator")
        result = simulator.run(circuit).result()
        statevector = result.get_statevector(circuit)
        prob_vector = Statevector(statevector).probabilities()
        return prob_vector
    else:
        raise NotImplementedError
\end{lstlisting}

\vspace{\baselineskip}
\begin{lstlisting}[caption={ Updated function to evaluate circuits supporting evaluation on GPUs. The target device for simulation is selected based on the `device' argument passed to the function. },
label={listing:impl/eval-gpu},
language=Python,
float]
def evaluate_circ(circuit, backend, device):
    if backend == "statevector_simulator":
        circuit.save_statevector()
        if device == "CPU":
            aercpu = AerSimulator(method='statevector')
            result = aercpu.run(circuit).result()
        if device == "GPU":
            aergpu = AerSimulator(method='statevector', device='GPU')
            result = aergpu.run(circuit, blocking_enable=True, blocking_qubits=30).result()
        
        statevector = result.get_statevector(circuit)
        prob_vector = Statevector(statevector).probabilities()

        return prob_vector
    else:
        raise NotImplementedError
\end{lstlisting}

As this module is executed in parallel, a function calculates the ideal number of processes to spawn. This function is also updated to return the number of required processes depending on the target device where the job would be executed i.e. CPU or GPU. Listing \ref{listing:impl/num-workers} shows the updated function with the updated part highlighted using comments. The number of workers for the target device set to CPU is calculated based on three factors: the total number of jobs to be executed, the total available memory and the total number of CPU cores on the machine. On the other hand, calculating this for the target device set to GPUs is not so straightforward. Multiple `num\_workers' values were used to benchmark the runtime of jobs on GPUs and it was observed that setting it to 32 (half of the available CPU cores on Perlmutter nodes) gave the best results.

\vspace{\baselineskip}
\begin{lstlisting}[caption={ Updated function to calculate the ideal number of processes to spawn for parallel job execution. The number of workers for CPUs is calculated whereas the number of workers for GPUs is set to 32 after benchmarking for different values. },
label={listing:impl/num-workers},
language=Python,
float]
def get_num_workers(num_jobs, ram_required_per_worker, device):
    if device == "CPU":
        ram_avail = psutil.virtual_memory().available / 1024**3
        ram_avail = ram_avail / 4 * 3
        num_cpus = int(os.cpu_count() / 4 * 3)
        num_workers = int(min(ram_avail / ram_required_per_worker, num_jobs, num_cpus))
    # Updated part of the function.
    # The device option is added to the original function.
    if device == "GPU":
        num_workers = 32

    return num_workers
\end{lstlisting}

\subsubsection{Builder and Verifier}

This module is responsible for reconstructing the probability distribution based on the subcircuit probabilities obtained from the Evaluator module and verifying (optional) whether the reconstructed probability lies within a certain error margin of the original probability distribution. This is where the CutQC framework calculates the Kronecker products for the subcircuit terms (refer to Figure \ref{fig:edge-decomposition}). To do this, the probability vectors are represented as `tensors' and the `Tensorflow' library is used to compute the Kronecker products of these `tensors'. The module also supports reconstructing the probability distribution of specified qubits in the scenario where storing the entire statevector is not feasible. There is no change made to this module as a part of this study.

\subsection{Benchmarking}
\label{sec:impl/benchmarks}

This section provides more information about the benchmark scripts and how they were executed. Refer to Section \ref{sec:benchmarks} for more details about the circuits used for benchmarking the two simulation methods. All benchmarks were performed on the Perlmutter \cite{perlmutter} system at NERSC (National Energy Research Scientific Computing Center) in Lawrence Berkeley National Lab.  

\begin{figure}[htbp]
    \centering
    \includegraphics[alt={Flow diagram representing how circuits are saved to disk using the QPY module in Qiskit.}, width=\linewidth]{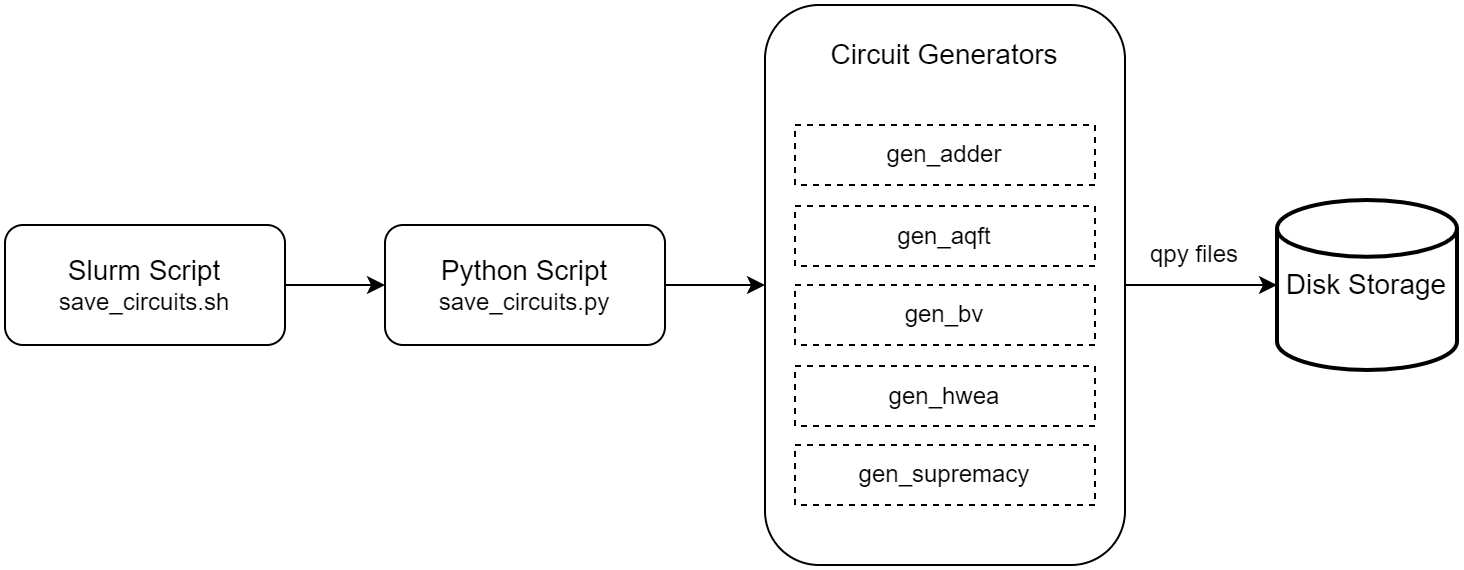}
    \caption{ This figure represents how the circuits are generated and saved as QPY files to avoid re-generating circuits for every benchmark run. A Slurm job runs a Python script which executes the required circuit generator and saves the circuit to disk using the \textit{QPY} module in Qiskit. }
    \label{fig:impl/flow-save-circ}
\end{figure}

Since the benchmarks reported in this study were executed multiple times, saving as much time as possible for actual computations was essential. To do this, circuits used in the benchmark were generated and saved beforehand. Actual benchmarks would then load these circuits from the disk and perform the necessary benchmarks. Figure \ref{fig:impl/flow-save-circ} shows the process of generating and saving the circuits to disk. A Slurm \cite{slurm} (workload manager software used by Perlmutter) job is used to execute the python script responsible for generating and saving quantum circuits to disk. The python script calls the circuit generator code which generates the required quantum circuit based on the input provided. This quantum circuit is then saved to disk in binary format using the `qpy' module in Qiskit.

The above process is repeated for all circuit combinations benchmarked (circuit type and number of qubits) in parallel. Benchmarks are executed once the quantum circuits are saved to disk. A Slurm job executes and manages the benchmark execution script for every circuit. Each circuit is passed on to the CutQC framework for evaluation using the circuit-cutting method. Finally, the circuit is simulated directly using the Qiskit Aer simulator and results for both simulation methods are saved to disk using the `pickle' module in Python.

\section{Methodology}
\label{chap:methodology}

This section aims to provide more information on the benchmarks performed in this study. Section \ref{sec:environment} talks about the hardware and software used for the benchmarks, and detailed information about the benchmarks is provided in Section \ref{sec:benchmarks}.

\subsection{Hardware and Software Environments}
\label{sec:environment}

\subsubsection{Hardware Environment}

All benchmarks were performed on Perlmutter \cite{perlmutter} (operated by NERSC at Lawrence Berkeley National Lab). Each benchmark was executed on a dedicated GPU node with several nodes being available for running multiple benchmarks at once.  

Each GPU node on Perlmutter has an AMD EPYC 7763 (Milan) CPU at 2.45 GHz, 256GB DDR4 RAM and four NVIDIA A100 GPUs \cite{perlmutter_specs}. There are two variants of these GPU nodes: a normal one with 40GB NVIDIA A100 GPUs and a high-bandwidth one with 80GB NVIDIA A100 GPUs. Benchmarks up to 30 qubits were executed on normal GPU nodes, whereas circuits with more than 30 qubits were executed on high-bandwidth GPU nodes.

\subsubsection{Software Environment}

All quantum circuit simulations were performed using the Qiskit Aer simulator \cite{qiskitPaper}\cite{qiskit}\cite{qiskitAerGPU} with GPUs on a server with SUSE Linux Enterprise Server 15 SP4. To split a quantum circuit into sub-circuits, the benchmarks use the CutQC framework \cite{cutqc} based on the work by Tang, W. et. al. \cite{tang2021cutqc}.  

\noindent Below is a list of all major software used in the benchmarks and their respective versions.
\begin{itemize}
    \item Python 3.12.2
    \item Qiskit 1.0.2
    \item Qiskit-Aer-GPU 0.14.0.1
\end{itemize}

\subsection{Benchmarks}
\label{sec:benchmarks}

The circuits used in the benchmarks are selected from the CutQC paper. The benchmarks were run for the number of qubits ranging from 10 to 34 (the maximum number of qubits we can simulate using one node of Perlmutter). Two methods are used to simulate the circuits: the first one simulates the circuit as it is (an uncut circuit), and the other uses CutQC to split the circuit into multiple subcircuits and simulate these subcircuits.  

The evaluation runtime, i.e., the time required to simulate all circuits using Qiskit Aer, is the main performance metric captured to compare the simulation methods. Apart from this, the runtime is also captured for different phases of the circuit-splitting method to gain more insights into the runtime scaling. Finally, circuit-splitting metadata (which includes the subcircuit width and number of effective qubits for every subcircuit generated) is also captured, which helps us to co-relate some of the observed trends in the runtime plots.

\vspace{\baselineskip}
\noindent More information about the circuits benchmarked is listed below.

\begin{enumerate}
    \item \textit{Adder}: This is a quantum ripple-carry addition circuit with a single ancilla qubit and linear depth based on the work of Cuccaro, S. A. et. al. \cite{cuccaro2004new}. It can be used to sum two quantum registers of the same width. It is benchmarked for an even number of qubits.
    \item \textit{Approximate Quantum Fourier Transform (AQFT)}: Work done by Barenco, A. et al. \cite{barenco1996approximate} proposes that as far as periodicity estimation is concerned, the AQFT algorithm yields better results than the QFT (Quantum Fourier Transform) algorithm. It is also benchmarked for an even number of qubits and a standard implementation from the Qiskit circuit library is used to generate the circuit.
    \item \textit{Berstein-Vazirani (BV)}: A popular quantum algorithm based on the work of Bernstein, E. et. al. \cite{bernstein1993quantum} which solves the hidden string problem. The implementation used in the benchmark is referenced from the QED-C \cite{qedcBenchmark} benchmark suite.
    \item \textit{Hardware efficient ansatz (HWEA)}: An example of quantum variational algorithms inspired by the work of Moll, N. et. al. \cite{moll2018quantum}.
    \item \textit{Supremacy}: A type of 2-D random circuit based on the work by Boixo, S. et. al. \cite{boixo2018characterizing} to characterize quantum supremacy. Similar to the CutQC paper, benchmarks are performed for near-square-shaped (e.g. 3*4) circuit layouts (which are harder to cut).
\end{enumerate}

\noindent All the circuits used in the benchmarks are not dense (having a huge circuit depth) quantum circuits but are rather sparse when it comes to the layout of the gates applied to the qubits. This is necessary because we impose a constraint on the CutQC framework to cut a given circuit within 10 cuts and generate a maximum of 5 subcircuits. While cutting dense circuits like Grover's (implemented without using ancilla qubits) is theoretically possible using the framework, the number of cuts required to cut such a circuit is very high, leading to enormous classical post-processing costs and is therefore not practical to simulate. One of the goals of these benchmarks is to study the impact of different sparse layouts of circuits on the runtime and the incurred classical post-processing cost when these circuits are simulated using \textit{circuit splitting}.
\section{Results}
\label{chap:results}

This section reports the results of the performed benchmarks. Before diving into the benchmarks, a few things need to be noted. The original CutQC paper uses a cluster of 16 nodes to perform the benchmarks. Regrettably, the multi-node version of the code is no longer accessible. Hence, the benchmarks in this study are conducted solely on a single node, restricting the simulation to a maximum of 34 qubits. This favors the full circuit simulation method since minimal overhead is present to simulate circuits on different GPUs on a single node. Full circuit executions on multiple nodes are subject to additional communication overhead (due to inter-node data exchanges required using MPI), which would have given a better idea of how the two simulation methods perform when performing simulations for a larger number of qubits (greater than 40).

\begin{figure*}[htbp]
\centering
\begin{subfigure}{0.33\textwidth}
  \centering
  \includegraphics[alt={Comparison of runtime for the Adder circuit using CutQC framework and Full circuit simulation}, width=\textwidth]{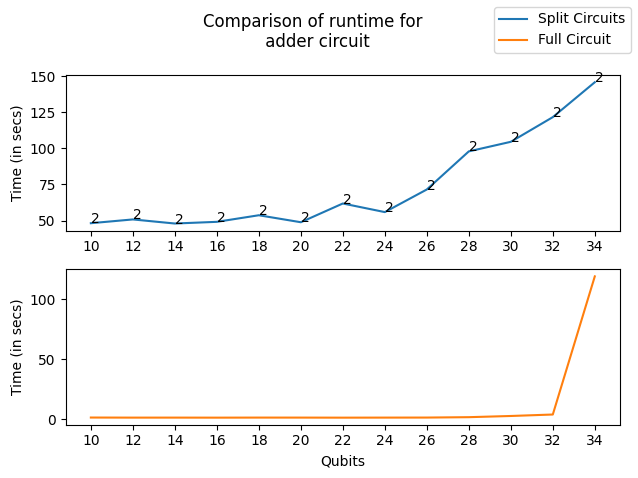}
  \caption{Adder}
  \label{fig:results/adder-annot}
\end{subfigure}%
\begin{subfigure}{0.33\textwidth}
  \centering
  \includegraphics[alt={Comparison of runtime for the AQFT circuit using CutQC framework and Full circuit simulation}, width=\textwidth]{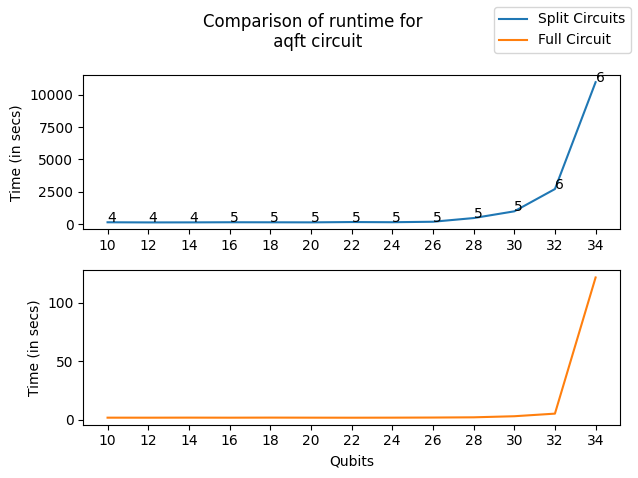}
  \caption{AQFT}
  \label{fig:results/aqft-annot}
\end{subfigure}%
\begin{subfigure}{0.33\textwidth}
  \centering
  \includegraphics[alt={Comparison of runtime for the BV circuit using CutQC framework and Full circuit simulation}, width=\textwidth]{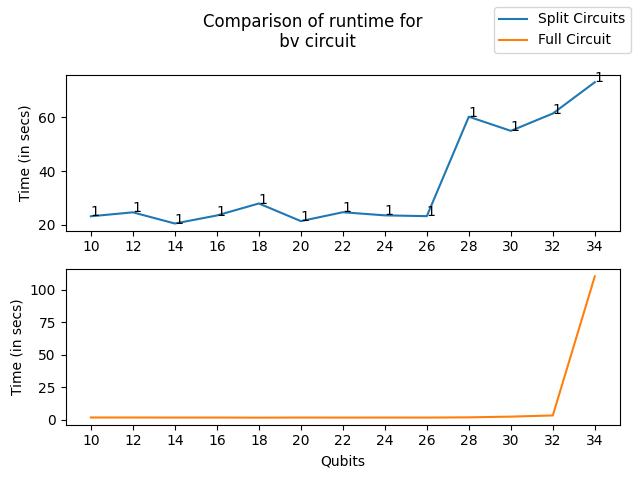}
  \caption{BV}
  \label{fig:results/bv-annot}
\end{subfigure}%

\begin{subfigure}{0.33\textwidth}
  \centering
  \includegraphics[alt={Comparison of runtime for the HWEA circuit using CutQC framework and Full circuit simulation}, width=\textwidth]{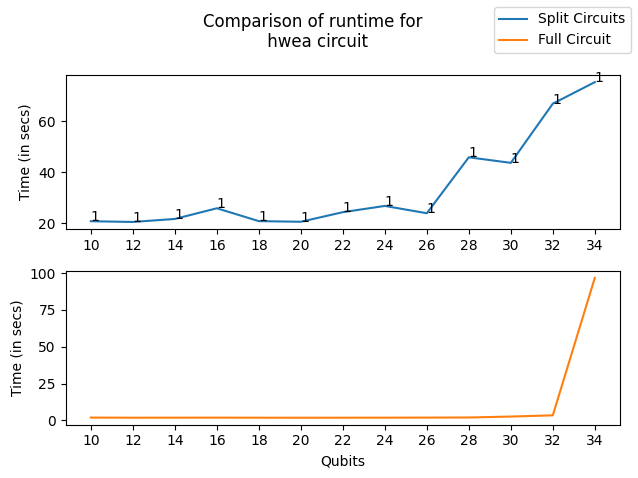}
  \caption{HWEA}
  \label{fig:results/hwea-annot}
\end{subfigure}%
\begin{subfigure}{0.33\textwidth}
  \centering
  \includegraphics[alt={Comparison of runtime for the Supremacy circuit using CutQC framework and Full circuit simulation}, width=\textwidth]{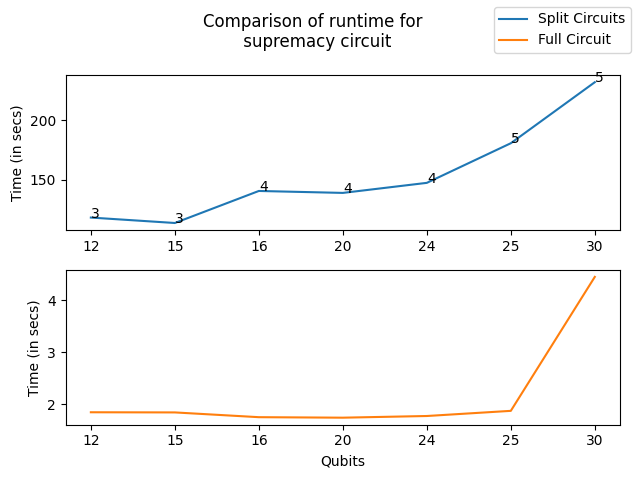}
  \caption{Supremacy}
  \label{fig:results/supremacy-annot}
\end{subfigure}
\caption{ Results for the runtime evaluations of all circuits. The top represents the evaluation time taken when the circuit is simulated by splitting into subcircuits, and the bottom represents the runtime for the uncut circuit. The top is further annotated to show the number of cuts made to split the original circuit. }
\label{fig:results}
\end{figure*}

\vspace{\baselineskip}
\vspace{\baselineskip}
Every circuit except the \textit{Supremacy} circuit is benchmarked for an even number of qubits ranging from 10 to 34. The \textit{Supremacy} circuit was benchmarked for 12, 15, 16, 20, 24, 25 and 30 qubits. Runtime evaluation results for all circuit types can be found in Figure \ref{fig:results}.

%
%
\vspace{\baselineskip}
Figure \ref{fig:results/adder-annot} shows the circuit evaluation times for circuit splitting versus full circuit execution for the \textit{Adder} circuit. As can be seen from the figure, the circuit splitting method runs close to two magnitudes slower than the full circuit execution for most of the qubit values. This is expected when running on a single node, as we run only one instance of the full circuit compared to the numerous instances of the subcircuits in the circuit-splitting method. When a circuit is cut into subcircuits using $K$ cuts, the number of subcircuits to simulate is in the order $4^K$ since we need to evaluate subcircuits with all possible initialization and measurements on qubit wires that are cut. The key thing to note here is that the runtime of the split circuits does not increase exponentially upon increasing the number of qubits. In fact, we can observe that the runtime remains the same for qubit values less than 26, after which there is a linear increase in the runtime. This is because all the \textit{Adder} circuits are split using two cuts, so the scaling of runtime in this case purely depends on the number of qubits being simulated. The same trend can also be observed for the \textit{BV} (Figure \ref{fig:results/bv-annot}) and \textit{HWEA} (Figure \ref{fig:results/hwea-annot}) circuits where all circuits are cut using the same number of cuts.  

A different trend can be observed for the \textit{AQFT} (Figure \ref{fig:results/aqft-annot}) and \textit{Supremacy} (Figure \ref{fig:results/supremacy-annot}) circuits where the number of cuts made increase as we increase the number of qubits. In the \textit{AQFT} circuit, the runtime increases more than 2x for the 32-qubit circuit compared to the 30-qubit circuit and an even sharper jump in runtime is observed for the 34-qubit circuit. This is because the 30-qubit circuit uses 5 cuts to split the circuits, whereas the 32 and 34-qubit circuits use 6 cuts to split the circuit, which results in exponentially more subcircuits to be simulated for the circuits having more than 30 qubits. The same trend is observed for the \textit{Supremacy} circuit when going from the 24-qubit circuit (using 4 cuts) to the circuits having more than 24 qubits (using 5 cuts).

%
%
\vspace{\baselineskip}
Another thing to notice from the circuit evaluation plots is the similarity between the runtime of the \textit{BV} and \textit{HWEA} circuits. The cause of this similarity lies in the way these two circuits are cut and split into subcircuits. Tables \ref{tab:results/subcirc-metadata-1} and \ref{tab:results/subcirc-metadata-2} contain the metadata for the split circuits for all the circuit types used in this study. Focusing on the metadata for the \textit{BV} and \textit{HWEA} circuits, it can be observed that both the circuits are split into two subcircuits with the same width and same number of effective qubits (qubits which contribute to the final uncut state). This result outlines an important piece of information that subcircuit evaluation runtime is closely related to how the cuts are made and the width of the resulting subcircuits. Another example of this can be seen by observing the metadata for the \textit{AQFT} circuit, where the difference between the overall width of subcircuits and the width of the uncut circuit is greater than the other circuit types. This is a direct consequence of requiring more cuts to split the original circuit as $ \textit{Width of original circuit} + \textit{Number of cuts} = \textit{Overall width of subcircuits} $.

\begin{table}[!htbp]
\resizebox{\linewidth}{!}{%
\begin{tabular}{|c|ll|ll|ll|ll|}
\hline
\multirow{3}{*}{\textbf{Qubits}} & \multicolumn{2}{c|}{\textbf{Adder}}                                                                                                                                                                            & \multicolumn{2}{c|}{\textbf{AQFT}}                                                                                                                                                                             & \multicolumn{2}{c|}{\textbf{BV}}                                                                                                                                                                               & \multicolumn{2}{c|}{\textbf{HWEA}}                                                                                                                                                                             \\ \cline{2-9} 
                                 & \multicolumn{1}{c|}{\multirow{2}{*}{\textbf{\begin{tabular}[c]{@{}c@{}}Subcirc.\\ width\end{tabular}}}} & \multicolumn{1}{c|}{\multirow{2}{*}{\textbf{\begin{tabular}[c]{@{}c@{}}Eff.\\ Qubits\end{tabular}}}} & \multicolumn{1}{c|}{\multirow{2}{*}{\textbf{\begin{tabular}[c]{@{}c@{}}Subcirc.\\ width\end{tabular}}}} & \multicolumn{1}{c|}{\multirow{2}{*}{\textbf{\begin{tabular}[c]{@{}c@{}}Eff.\\ Qubits\end{tabular}}}} & \multicolumn{1}{c|}{\multirow{2}{*}{\textbf{\begin{tabular}[c]{@{}c@{}}Subcirc.\\ width\end{tabular}}}} & \multicolumn{1}{c|}{\multirow{2}{*}{\textbf{\begin{tabular}[c]{@{}c@{}}Eff.\\ Qubits\end{tabular}}}} & \multicolumn{1}{c|}{\multirow{2}{*}{\textbf{\begin{tabular}[c]{@{}c@{}}Subcirc.\\ width\end{tabular}}}} & \multicolumn{1}{c|}{\multirow{2}{*}{\textbf{\begin{tabular}[c]{@{}c@{}}Eff.\\ Qubits\end{tabular}}}} \\
                                 & \multicolumn{1}{c|}{}                                                                                   & \multicolumn{1}{c|}{}                                                                                & \multicolumn{1}{c|}{}                                                                                   & \multicolumn{1}{c|}{}                                                                                & \multicolumn{1}{c|}{}                                                                                   & \multicolumn{1}{c|}{}                                                                                & \multicolumn{1}{c|}{}                                                                                   & \multicolumn{1}{c|}{}                                                                                \\ \hline
\multirow{2}{*}{\textbf{10}}     & \multicolumn{1}{l|}{6}                                                                                  & 5                                                                                                    & \multicolumn{1}{l|}{8}                                                                                  & 4                                                                                                    & \multicolumn{1}{l|}{6}                                                                                  & 5                                                                                                    & \multicolumn{1}{l|}{6}                                                                                  & 5                                                                                                    \\ \cline{2-9} 
                                 & \multicolumn{1}{l|}{6}                                                                                  & 5                                                                                                    & \multicolumn{1}{l|}{6}                                                                                  & 6                                                                                                    & \multicolumn{1}{l|}{5}                                                                                  & 5                                                                                                    & \multicolumn{1}{l|}{5}                                                                                  & 5                                                                                                    \\ \hline
\multirow{2}{*}{\textbf{12}}     & \multicolumn{1}{l|}{6}                                                                                  & 5                                                                                                    & \multicolumn{1}{l|}{8}                                                                                  & 4                                                                                                    & \multicolumn{1}{l|}{7}                                                                                  & 6                                                                                                    & \multicolumn{1}{l|}{7}                                                                                  & 6                                                                                                    \\ \cline{2-9} 
                                 & \multicolumn{1}{l|}{8}                                                                                  & 7                                                                                                    & \multicolumn{1}{l|}{8}                                                                                  & 8                                                                                                    & \multicolumn{1}{l|}{6}                                                                                  & 6                                                                                                    & \multicolumn{1}{l|}{6}                                                                                  & 6                                                                                                    \\ \hline
\multirow{2}{*}{\textbf{14}}     & \multicolumn{1}{l|}{8}                                                                                  & 7                                                                                                    & \multicolumn{1}{l|}{8}                                                                                  & 4                                                                                                    & \multicolumn{1}{l|}{6}                                                                                  & 5                                                                                                    & \multicolumn{1}{l|}{6}                                                                                  & 5                                                                                                    \\ \cline{2-9} 
                                 & \multicolumn{1}{l|}{8}                                                                                  & 7                                                                                                    & \multicolumn{1}{l|}{10}                                                                                 & 10                                                                                                   & \multicolumn{1}{l|}{9}                                                                                  & 9                                                                                                    & \multicolumn{1}{l|}{9}                                                                                  & 9                                                                                                    \\ \hline
\multirow{2}{*}{\textbf{16}}     & \multicolumn{1}{l|}{8}                                                                                  & 7                                                                                                    & \multicolumn{1}{l|}{12}                                                                                 & 7                                                                                                    & \multicolumn{1}{l|}{10}                                                                                 & 9                                                                                                    & \multicolumn{1}{l|}{10}                                                                                 & 9                                                                                                    \\ \cline{2-9} 
                                 & \multicolumn{1}{l|}{10}                                                                                 & 9                                                                                                    & \multicolumn{1}{l|}{9}                                                                                  & 9                                                                                                    & \multicolumn{1}{l|}{7}                                                                                  & 7                                                                                                    & \multicolumn{1}{l|}{7}                                                                                  & 7                                                                                                    \\ \hline
\multirow{2}{*}{\textbf{18}}     & \multicolumn{1}{l|}{8}                                                                                  & 7                                                                                                    & \multicolumn{1}{l|}{10}                                                                                 & 5                                                                                                    & \multicolumn{1}{l|}{11}                                                                                 & 10                                                                                                   & \multicolumn{1}{l|}{11}                                                                                 & 10                                                                                                   \\ \cline{2-9} 
                                 & \multicolumn{1}{l|}{12}                                                                                 & 11                                                                                                   & \multicolumn{1}{l|}{13}                                                                                 & 13                                                                                                   & \multicolumn{1}{l|}{8}                                                                                  & 8                                                                                                    & \multicolumn{1}{l|}{8}                                                                                  & 8                                                                                                    \\ \hline
\multirow{2}{*}{\textbf{20}}     & \multicolumn{1}{l|}{10}                                                                                 & 9                                                                                                    & \multicolumn{1}{l|}{14}                                                                                 & 9                                                                                                    & \multicolumn{1}{l|}{13}                                                                                 & 12                                                                                                   & \multicolumn{1}{l|}{13}                                                                                 & 12                                                                                                   \\ \cline{2-9} 
                                 & \multicolumn{1}{l|}{12}                                                                                 & 11                                                                                                   & \multicolumn{1}{l|}{11}                                                                                 & 11                                                                                                   & \multicolumn{1}{l|}{8}                                                                                  & 8                                                                                                    & \multicolumn{1}{l|}{8}                                                                                  & 8                                                                                                    \\ \hline
\multirow{2}{*}{\textbf{22}}     & \multicolumn{1}{l|}{10}                                                                                 & 9                                                                                                    & \multicolumn{1}{l|}{16}                                                                                 & 11                                                                                                   & \multicolumn{1}{l|}{13}                                                                                 & 12                                                                                                   & \multicolumn{1}{l|}{13}                                                                                 & 12                                                                                                   \\ \cline{2-9} 
                                 & \multicolumn{1}{l|}{14}                                                                                 & 13                                                                                                   & \multicolumn{1}{l|}{11}                                                                                 & 11                                                                                                   & \multicolumn{1}{l|}{10}                                                                                 & 10                                                                                                   & \multicolumn{1}{l|}{10}                                                                                 & 10                                                                                                   \\ \hline
\multirow{2}{*}{\textbf{24}}     & \multicolumn{1}{l|}{10}                                                                                 & 9                                                                                                    & \multicolumn{1}{l|}{17}                                                                                 & 12                                                                                                   & \multicolumn{1}{l|}{15}                                                                                 & 14                                                                                                   & \multicolumn{1}{l|}{15}                                                                                 & 14                                                                                                   \\ \cline{2-9} 
                                 & \multicolumn{1}{l|}{16}                                                                                 & 15                                                                                                   & \multicolumn{1}{l|}{12}                                                                                 & 12                                                                                                   & \multicolumn{1}{l|}{10}                                                                                 & 10                                                                                                   & \multicolumn{1}{l|}{10}                                                                                 & 10                                                                                                   \\ \hline
\multirow{2}{*}{\textbf{26}}     & \multicolumn{1}{l|}{16}                                                                                 & 15                                                                                                   & \multicolumn{1}{l|}{18}                                                                                 & 13                                                                                                   & \multicolumn{1}{l|}{17}                                                                                 & 16                                                                                                   & \multicolumn{1}{l|}{17}                                                                                 & 16                                                                                                   \\ \cline{2-9} 
                                 & \multicolumn{1}{l|}{12}                                                                                 & 11                                                                                                   & \multicolumn{1}{l|}{13}                                                                                 & 13                                                                                                   & \multicolumn{1}{l|}{10}                                                                                 & 10                                                                                                   & \multicolumn{1}{l|}{10}                                                                                 & 10                                                                                                   \\ \hline
\multirow{2}{*}{\textbf{28}}     & \multicolumn{1}{l|}{14}                                                                                 & 13                                                                                                   & \multicolumn{1}{l|}{19}                                                                                 & 14                                                                                                   & \multicolumn{1}{l|}{18}                                                                                 & 17                                                                                                   & \multicolumn{1}{l|}{18}                                                                                 & 17                                                                                                   \\ \cline{2-9} 
                                 & \multicolumn{1}{l|}{16}                                                                                 & 15                                                                                                   & \multicolumn{1}{l|}{14}                                                                                 & 14                                                                                                   & \multicolumn{1}{l|}{11}                                                                                 & 11                                                                                                   & \multicolumn{1}{l|}{11}                                                                                 & 11                                                                                                   \\ \hline
\multirow{2}{*}{\textbf{30}}     & \multicolumn{1}{l|}{12}                                                                                 & 11                                                                                                   & \multicolumn{1}{l|}{21}                                                                                 & 16                                                                                                   & \multicolumn{1}{l|}{19}                                                                                 & 18                                                                                                   & \multicolumn{1}{l|}{19}                                                                                 & 18                                                                                                   \\ \cline{2-9} 
                                 & \multicolumn{1}{l|}{20}                                                                                 & 19                                                                                                   & \multicolumn{1}{l|}{14}                                                                                 & 14                                                                                                   & \multicolumn{1}{l|}{12}                                                                                 & 12                                                                                                   & \multicolumn{1}{l|}{12}                                                                                 & 12                                                                                                   \\ \hline
\multirow{2}{*}{\textbf{32}}     & \multicolumn{1}{l|}{14}                                                                                 & 13                                                                                                   & \multicolumn{1}{l|}{22}                                                                                 & 16                                                                                                   & \multicolumn{1}{l|}{21}                                                                                 & 20                                                                                                   & \multicolumn{1}{l|}{21}                                                                                 & 20                                                                                                   \\ \cline{2-9} 
                                 & \multicolumn{1}{l|}{20}                                                                                 & 19                                                                                                   & \multicolumn{1}{l|}{16}                                                                                 & 16                                                                                                   & \multicolumn{1}{l|}{12}                                                                                 & 12                                                                                                   & \multicolumn{1}{l|}{12}                                                                                 & 12                                                                                                   \\ \hline
\multirow{2}{*}{\textbf{34}}     & \multicolumn{1}{l|}{14}                                                                                 & 13                                                                                                   & \multicolumn{1}{l|}{24}                                                                                 & 18                                                                                                   & \multicolumn{1}{l|}{22}                                                                                 & 21                                                                                                   & \multicolumn{1}{l|}{22}                                                                                 & 21                                                                                                   \\ \cline{2-9} 
                                 & \multicolumn{1}{l|}{22}                                                                                 & 21                                                                                                   & \multicolumn{1}{l|}{16}                                                                                 & 16                                                                                                   & \multicolumn{1}{l|}{13}                                                                                 & 13                                                                                                   & \multicolumn{1}{l|}{13}                                                                                 & 13                                                                                                   \\ \hline
\end{tabular}%
}
\caption{This table contains the metadata of the split circuits for \textit{Adder}, \textit{AQFT}, \textit{BV} and \textit{HWEA} circuit types. Subcircuit width and the Effective qubits (which contribute to the original uncut state) of that subcircuit are reported for every circuit benchmarked. In this benchmark, every circuit is split into two subcircuits, which is why there are only two rows for every circuit benchmarked.}
\label{tab:results/subcirc-metadata-1}
\end{table}

\begin{table}[!htbp]
\resizebox{\linewidth}{!}{%
\begin{tabular}{|ccl|ll|ll|ll|ll|ll|ll|ll|}
\hline
\multicolumn{3}{|c|}{\textbf{Qubits}}                                                                                                              & \multicolumn{2}{c|}{\textbf{12}} & \multicolumn{2}{c|}{\textbf{15}} & \multicolumn{2}{c|}{\textbf{16}} & \multicolumn{2}{c|}{\textbf{20}} & \multicolumn{2}{c|}{\textbf{24}} & \multicolumn{2}{c|}{\textbf{25}} & \multicolumn{2}{c|}{\textbf{30}} \\ \hline
\multicolumn{1}{|c|}{\multirow{2}{*}{\textbf{Supremacy}}} & \multicolumn{2}{c|}{\textbf{\begin{tabular}[c]{@{}c@{}}Subcirc.\\ width\end{tabular}}} & \multicolumn{1}{l|}{8}    & 7    & \multicolumn{1}{l|}{10}    & 8   & \multicolumn{1}{l|}{12}    & 8   & \multicolumn{1}{l|}{11}   & 13   & \multicolumn{1}{l|}{11}   & 17   & \multicolumn{1}{l|}{18}   & 12   & \multicolumn{1}{l|}{19}   & 16   \\ \cline{2-17} 
\multicolumn{1}{|c|}{}                                    & \multicolumn{2}{c|}{\textbf{\begin{tabular}[c]{@{}c@{}}Eff.\\ Qubits\end{tabular}}}    & \multicolumn{1}{l|}{7}    & 5    & \multicolumn{1}{l|}{7}     & 8   & \multicolumn{1}{l|}{9}     & 7   & \multicolumn{1}{l|}{10}   & 10   & \multicolumn{1}{l|}{8}    & 16   & \multicolumn{1}{l|}{14}   & 11   & \multicolumn{1}{l|}{18}   & 12   \\ \hline
\end{tabular}%
}
\caption{This table contains the metadata of the split circuits for \textit{Supremacy} circuit type. Subcircuit width and the Effective qubits (which contribute to the original uncut state) of each subcircuit are reported for every circuit benchmarked. In this benchmark, every circuit is split into two subcircuits, which is why there are only two rows for every circuit benchmarked.}
\label{tab:results/subcirc-metadata-2}
\end{table}

\begin{table}[!htbp]
\resizebox{\linewidth}{!}{%
\begin{tabular}{|c|c|}
\hline
\textbf{Circuit Width} & \textbf{\begin{tabular}[c]{@{}c@{}}Approximate Memory Required to store Statevector\\ (in GB)\end{tabular}} \\ \hline
10                     & $1.63 \times 10^{-5}$                                                                                                   \\ \hline
12                     & $6.55 \times 10^{-5}$                                                                                                   \\ \hline
14                     & $2.62 \times 10^{-4}$                                                                                                    \\ \hline
15                     & $5.24 \times 10^{-4}$                                                                                                    \\ \hline
16                     & $1.04 \times 10^{-3}$                                                                                                    \\ \hline
18                     & $4.19 \times 10^{-3}$                                                                                                   \\ \hline
20                     & $1.67 \times 10^{-2}$                                                                                                    \\ \hline
22                     & $6.71 \times 10^{-2}$                                                                                                    \\ \hline
24                     & 0.26                                                                                                        \\ \hline
25                     & 0.53                                                                                                        \\ \hline
26                     & 1.07                                                                                                        \\ \hline
28                     & 4.29                                                                                                        \\ \hline
30                     & 17.17                                                                                                       \\ \hline
32                     & 68.71                                                                                                       \\ \hline
34                     & 274.87                                                                                                      \\ \hline
\end{tabular}
}
\caption{Approximate Memory space required to store a Quantum Circuit in Gigabytes}
\label{tab:results/req_storage}
\end{table}

\subsection*{Findings and Discussion}

The benchmarking results show that a full-circuit simulation runs significantly faster than a circuit-splitting simulation for single-node executions. The results provide key insights into the performance of the circuit-splitting simulation method. Parallel execution of subcircuits lets us hide the exponential cost of simulation when the input circuit is cut using less than 6 cuts (at least for the type of circuits benchmarked in this study). Another key insight is that circuits cut similarly (i.e. having similar widths and effective qubits) have similar runtime trends. This hints that the runtime is agnostic of the layout of the gates present in the original circuit and that the runtime can be estimated based on the cut locations, the resulting subcircuit widths and the number of effective qubits. Tables \ref{tab:results/subcirc-metadata-1} and \ref{tab:results/subcirc-metadata-2} show the subcircuit widths after every circuit is cut. It can be observed that the width of the subcircuits is 30-40\% less than the original circuit. This enables us to simulate these circuits using significantly fewer resources than the full-circuit simulation. A 30-40\% reduction in circuit width significantly reduces the amount of resources required to simulate the circuit, as for every qubit reduced, the computational space is reduced by half (Refer Table \ref{tab:results/req_storage}). Considering all factors, simulating large circuits using circuit-splitting and full-circuit simulation methods is a tradeoff between runtime and memory requirements. A large circuit can be simulated faster if there are enormous resources available to simulate it. In contrast, if resources are limited, the circuit-splitting method can be used to simulate the required circuit. In an ideal scenario, we can leverage both these approaches to efficiently simulate large quantum circuits. Large circuits can be split to reduce the resources required to simulate the circuit, and on the other hand, the cuts can be minimized as much as possible to avoid incurring exponential runtime to simulate all subcircuits.
\section{Conclusions}
\label{chap:conclusions}

It is evident from the benchmark results that the Full circuit execution method greatly outperforms the Circuit Splitting simulation method. A major contributing factor is the exponential number of subcircuits to be evaluated depending on the number of cuts made to the original circuit. The parallel evaluation lets us hide this exponential nature but is effective only for circuits split using a small number of cuts. Circuits requiring more than a few cuts still take exponential time to be evaluated.  

An interesting benchmark for future work is evaluating circuits having a width of more than 40 qubits. Full circuit simulations for such circuits would require hundreds of nodes resulting in a considerable amount of MPI communication overhead to exchange data between these nodes. On the other hand, the Circuit Splitting evaluation method would enjoy the benefit of running small circuits capable of being evaluated on a single node. This would give us a better picture of how both the simulation methods work on a larger scale and whether any of these methods is feasible to use for large-scale but efficient quantum circuit simulations.  

A key thing to keep in mind is that the performance of the Circuit Splitting method relies on the efficient cutting of the circuit using a small number of cuts. However, this is not always possible for dense circuits or circuits having two-qubit gates acting on all pairs of qubits (e.g. Grover's algorithm circuit implemented without using ancilla qubits). These are scenarios where circuit-splitting algorithms leveraging gate-cutting can be beneficial. With new circuit-splitting algorithms being published every year, it is only a matter of time before we will be able to classically simulate circuits with a width of more than 100 qubits (albeit with a huge resource cost). This achievement will be a great boost for the scientific community to research and validate new quantum algorithms.

\section*{Acknowledgements}
\label{sec:acknowledgements}

This research used resources of the National Energy Research Scientific Computing Center (NERSC), a Department of Energy Office of Science User Facility using NERSC award DDR-ERCAP0026767.

\printbibliography


\end{document}